\documentclass[letter,scriptaddress,twocolumn, showkeys]{revtex4}
\usepackage{ amssymb }
	\usepackage{amsmath}
	\usepackage{makeidx}
	\usepackage{amsfonts}
	\usepackage{amsthm}

	\usepackage[ansinew]{inputenc}
	\usepackage[usenames,dvipsnames]{pstricks}
	\usepackage{subfigure}
	\usepackage{epsfig}
	\usepackage{pst-grad} 
	\usepackage{pst-plot} 
	\usepackage[colorlinks,hyperindex]{hyperref}
	\hypersetup
	{
		colorlinks,%
		citecolor=blue,%
		linkcolor=blue,%
		urlcolor=blue,%
	}
	\DeclareRobustCommand{\stirling}{\genfrac\{\}{0pt}{}}

\def\N{\mathbb{N}}

\def\I{\mathcal{I}}

\def\T{\mathcal{T}}

\def\<{\langle}
\def\>{\rangle}

\def\0{\mathbf{0}}

	\newtheorem{defn}{Definition}
	\newtheorem{example}{Example}

\makeindex

\begin{document}

\title{A trust model for spreading gossip in social networks}

\author{ Rinni Bhansali $^{a}$ and   Laura P. Schaposnik $^{b,c}$}

  \affiliation {(a)  Half Hollow Hills High School East, 50 Vanderbilt Pkwy, Dix Hills, NY 11746, USA.  \\
  (b)  University of Illinois at Chicago, Chicago, IL 60607, USA.\\
  (c) Simons Center for Geometry and Physics, NY 11794, USA.}

\begin{abstract}

We introduce here a multi-type bootstrap percolation model, which we call {\it $\T$-Bootstrap Percolation} ($\T$-BP), and apply it to study information propagation in social networks. In this model, a social network is represented by a graph $G$ whose vertices have different labels corresponding to the type of role the person plays in the network (e.g. a student, an educator, etc.). 
Once an initial set of vertices of $G$ is randomly selected to be carrying a gossip (e.g.~to be infected), the gossip propagates to a new vertex provided it is transmitted by a minimum threshold of vertices with different labels.  By considering  random graphs, which have been  shown to closely represent  social networks,  we study different properties of the $\T$-BP  model through  numerical simulations, and describe its implications when applied to rumour spread, fake news, and marketing strategies.

\end{abstract}

\maketitle
 

\section{Introduction}
\label{Introduction}
Most people have struggled at some point to find the perfect present for their beloved: we hear from our son's friends that certain ``bacteria growing kit'' would be   fun for his 10th birthday - but will it actually be safe? Once we hear from our son's friends' parents that the  ``bacteria growing kit''  is indeed entertaining and safe for that age, we are close to decided on buying it.  Is this recurrent phenomenon a consequence of a natural instinct that one has, where having the same information transmitted by different ``types'' of people inspires more trust? If so, we  naturally wonder:
 \begin{center}
 {\it How many different types of people (colleagues, friends, taxi drivers, etc.) should we hear a piece of informayion from, before we start transmitting it as a true fact?}
 \end{center}
  Or equivalently, and concerning marketing strategies, 
 \begin{center}
 {\it How many different types of people should recommend to us a service or a product before we  buy it, and begin recommending it ourselves?} \end{center}
 
Having a clear range for sources of information would allow members of the society to disbelieve gossips and differentiate fake news. Moreover, understanding this range would also allow the industry to target wisely a minimum amount of consumers within each type of people, and using the natural propagating process to continue the marketing on its own. 

In this paper, we build a new model of information/disease spread which we then use to understand the above questions. Our model builds upon the classical Bootstrap Percolation introduced in \cite{5}, but incorporates the concept of different types of members of society. 
Bootstrap percolation is a particular class of monotone cellular automata describing an   activation process which follows certain activation rules, and which has been much used to model interactions within societies. In particular, in the classical $r$-neighbour bootstrap process  on a graph $G$, a set $A$ of initially ``infected'' vertices spreads by infecting vertices with at least $r$ already-infected neighbours (e.g., see \cite{1,2,3,4,6,7}).

 \begin{figure}[h]
\centering
\resizebox{0.45\textwidth}{!}{%
  \includegraphics{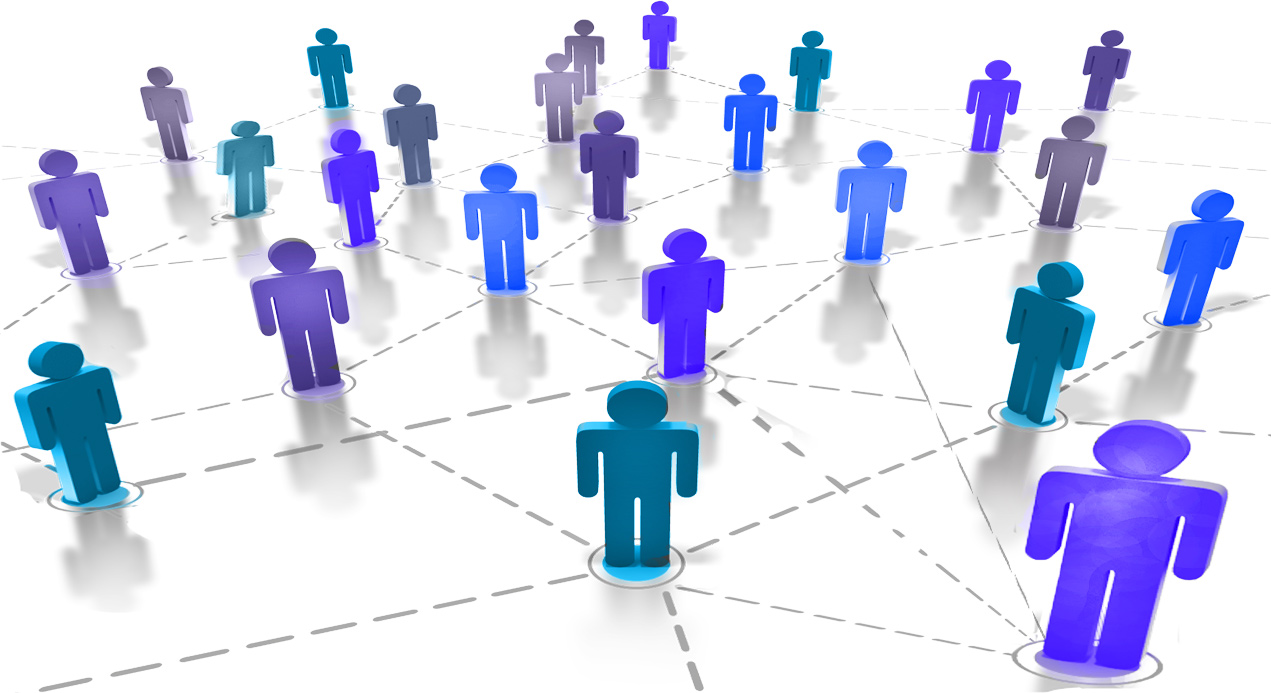}   
  }
\caption{Examples of  a social network where different colours - labels - describe the  types of people within gossip spread.}\label{Figure 1}
\end{figure}

In the present manuscript we shall introduce a multi-type version of bootstrap percolation which we call a \textit{\textbf{Trusted Bootstrap Percolation}} or \textit{\textbf{$\T$-Bootstrap Percolation}} ($\T$-BP)  to answer an equivalent question to those posed above:

 \begin{center}{\it How does information percolate when a messenger only passes the information if it has been received by a number of different sources of certain types?}\end{center}

In what follows, we use $\T$-BP to numerically understand the spread of gossip in random graphs $G$ simulating social networks. The simplest form of $\T$-BP is:   
\begin{defn}
Consider a finite or infinite graph $G$, two natural numbers $r,m\in \N$,  a vector $k=(k_1,\ldots,k_m)\in \N^m$ of non-negative numbers for which exactly $r$ of them satisfy $k_i\neq 0$,  and a set $A:=A_0$ of   initially ``infected'' vertices in $G$. After assigning randomly a label in $\{1,\ldots,m\}$ to each vertex, we define $r$-bootstrap percolation with trust level $k$ on $G$ as the process in which at each time step all of the vertices which have at least $k_i$ adjacent vertices infected with label $i$ become infected. \end{defn}
In what follows we shall introduce and study the most generic form of $\T$-BP in Sections \ref{intro}-\ref{mas}, and conclude the paper describing the implications of our results within society in Section \ref{last}.
\pagebreak
\section{Multi type Bootstrap Percolation}\label{intro}
In the simplest form of  $\T$-BP described above, at each time $t\in \N$, the set of infected vertices is given by
 \begin{eqnarray}A_{t+1} = A_t \cup \{v \in V(G) ~:~  |N_i(v) \cap A_t| \geq k_i ~\nonumber \\{\rm for~ ~}i=1,\ldots,m\},\end{eqnarray}
where $N_i(v)$ denotes the set of  adjacent vertices to $v$ in $G$ with label $i$. The most generic form of our multi-type bootstrap   percolation model is inspired by the concept of an update family $\mathcal{U}$ from \cite{Upaper}.

\begin{defn}[{\bf $\mathcal{T}$-bootstrap percolation}]
A {\it trust family} $\mathcal{T}:=\{K^1,\ldots,K^n\}$ is a tuple composed of trust vectors $K^j=(k_1^j,\ldots,k_m^j)\in \mathbb{N}^m$. Then, we define  \textbf{\textit{Trusted Bootstrap Percolation}}, or \textbf{\textit{ $\mathcal{T}$-bootstrap percolation}},  as the percolation process for which \linebreak $A_0:=A$ and at time $t+1$ the infected vertices are 
\begin{eqnarray}A_{t+1} = A_t \cup \{v \in V(G) ~:~\exists K_j\in \mathcal{T} {~s.t.}~\\  |N_i(v) \cap A_t| \geq k^j_i ~{\rm for~ ~}i=1,\ldots,m\}.\end{eqnarray}
\end{defn}
It is important to note that classical $r$-neighbour bootstrap percolation is a particular case of  $\mathcal{T}$-bootstrap percolation. 

\begin{example} An example of $r$-neighbour bootstrap is given by $\mathcal{T}$-bootstrap percolation for the trust family \begin{eqnarray}\mathcal{T}:=\left\{ K^j\in \mathbb{N}^m~\left|~\sum_{i=1}^{m}k_i^j=r\right.\right\}.
\end{eqnarray}Specifically, to recover $r$-neighbour bootstrap percolation with only one label one may consider $m=1$, and set $\mathcal{T}=\{K^1=(r)\}$. \label{rbp}
\end{example}
 When studying  $\T$-BP, it is useful to bear in mind its application to society. For this, in its simplest form, the above set up  of $\T$-BP corresponds to considering a society with $m$ different types of people, and a gossip that spreads only if it's passed by $k_j$ number of people of type $j$. In the most generic set up, the requirement for a gossip to spread is given by the existence of at least one trust vector $K^i$ for which the gossip can be passed by $k_j$ people of type $j$, for all types $j$. 
 For ease of notation, we shall refer to \textit{\textbf{ $r$-neighbour $\T$-BP}} when considering a $\T$-BP model with exactly $r$ integers  $k^j\in \{0,1\}$ non-zero. 
 
\begin{example}Consider $\T$-BP   with $m=3,~r=2$ and $k_i=1$. In this case, the model can be used to represent the spread of a political rumour among a society of Democrats, Republicans, and Independents. Labelling the vertices with $1,2,3$ to represent each political party,  suppose that, in order to limit the spread of biased (and potentially false) information, there exists a rule that an individual will only believe and pass on the rumour, if he/she heard it from 2 people with different political backgrounds. Then, the trust family in this model would be $\T = \{(1,1,0),(1,0,1),(0,1,1)\}$, and this is equivalent  $2$-neighbour $\T$-BP with 3 labels.\label{demo}
\end{example}



\section{Immunity to gossip spread}
In the following subsections we shall focus on two forms of $\mathcal{T}$-bootstrap percolation that are of particular interest: $r$-neighbour $\T$-BP, and the simplest $\T$-BP, which has single trust vector $k$.
Within the $\T$-BP a proportion of the vertices is {\it immune} to the infection: these vertices do not belong to $A_t$ for any $t\in \mathbb{N}$, and we shall formally define the   \textit{immune set} by   \begin{eqnarray}
\I:=\left\{  v\in G ~:~ \forall   K^j\in \mathcal{T}, ~ \exists i\in[m]~s.t.~|N_i(v)|<k_i^j  \right\},\nonumber
\end{eqnarray}
    for $\T = \{K^1, ..., K^n\}$ the trust family, and $K^i = \{k_1^j, ..., k_m^j\} \in \mathbb{N}^m$ trust vectors as before.
   
In order to study the $\T$-BP on $G$, the vertices in $\I$ need to be removed from $G$, leading  to a modified graph which sometimes will be disconnected, with some components that might never become infected.
To understand the likelihood of percolation of a $\T$-BP model and  how immune vertices disrupt percolation on a network, it is useful to introduce the notion of {\it diversity}: 
given a vertex $v\in G$ in a $\T$-BP model, define the {\it diversity of $v$} as the number $D_v$ of different labels that vertices  have: 
\begin{eqnarray}
D_v:=\left| \{ i\in \{1,\ldots,m\}~:~ N_i(v)\neq \emptyset \} \right |. 
\end{eqnarray} 
 
 \subsection{Immunity for $r$-neighbour $\T$-BP}
In the case of  $r$-neighbour $\T$-BP, a vertex is   {\it immune} if it does not have neighbours of at least $r$ distinct labels.  
 From the definition of $r$-neighbour $\T$-BP, one can see that if a vertex is not immune, then  $D_v\geq r$.  Hence, it is of particular interest to understand which vertices $v$ have $D_v<r$, since those will comprise all of the immune vertices. 
 \begin{example}Returning to Example \ref{demo}, if an individual knows only Democrats, then it is impossible for them to be infected with gossip.\end{example}

 Consider $\T$-BP on a graph $G$, and let each vector component $k_j\in \{0,1\}$. Then, the probability   that a vertex $v$ with $|N(v)|=d$ has $D_v\leq r-1$   is \begin{eqnarray}
 P(D_v\leq r-1)=  \frac{1}{m^d} \sum\limits_{j = 1}^{r-1} \left[{m\choose j}(j!)\stirling{d}{j}\right]
,
 \end{eqnarray}
where $\stirling{d}{j}$ is the Sterling number of the second kind (the reader may refer to Appendix \ref{AA} for the proof of this statement).
In particular, when the number of different labels required is equal to the number of labels that exist, i.e., when $m=r$, then 
 \begin{eqnarray} P(D_v\leq r-1)=  \frac{1}{r^d} \sum\limits_{j = 1}^{r-1} \left[\frac{r!}{(r-j)!}\stirling{d}{j}\right].
 \end{eqnarray}

\pagebreak
 \subsection{Immunity for the simplest form of $\T$-BP}

The second form of $\T$-bootstrap percolation we shall consider is the simplest form of $\T$-BP in which there is a single trust vector $k = (k_1, k_2, ..., k_m)$. In this setting, given a vertex $v$ with degree $d$ on a graph $G$, consider the vector  $x = (x_1, x_2, ..., x_m)$ where $x_i = |N_i(v)|$ is the number of adjacent vertices with label $i$.  In particular,  $\sum_{i=1}^m x_i  = d$. Moreover, note that a vertex is immune if  $x_i < k_i$  for some integer $i$. In this case, the probability of immunity  $p^d_I(k)$ for a vertex $v$ with $|N(v)|=d$ is 
 
\begin{eqnarray}
   &~& p^d_I(k) =
        \notag 1- \sum \limits_{x_m = k_m}^{d-(\sum_{l=1}^{m-1}k_l)}{d\choose x_m}\left(\frac{1}{m}\right)^{x_m} \\
        \notag   &\bigg[&\sum \limits_{x_{m-1} = k_{m-1}}^{d-x_m-(\sum_{l=1}^{m-1}k_l)}{d-x_m\choose x_{m-1}}\left(\frac{1}{m}\right)^{x_{m-2}} \bigg[ \cdots \\ 
        \notag  &\bigg[&\sum \limits_{x_2 = k_2}^{x_1+x_2-k_1}{x_1+x_2\choose x_2}\left(\frac{1}{m}\right)^{x_2}\left(\frac{1}{m}\right)^{x_1)} \bigg] \bigg] ... \bigg]
\label{second}
\end{eqnarray}

\noindent The reader should refer to Appendix \ref{BB} for the proof of the above equation. 
 
Given a $\T$-BP on a graph $G$, we shall denote by $p_I(G,\T)$  the expected fraction of immune vertices on $G$, and by $p^d_I(\T)$   the probability of immunity for a vertex of degree $d$  with trust family $\T$. Then, one can show that 
in  a $\T$-BP on a graph $G$ with degree distribution $P(d)$, one has that  the fraction of immune vertices is \begin{eqnarray}
     p_I(G,\T)=\sum \limits_{j=0}^{\infty} p^j_I(\T)P(j).
\end{eqnarray}

 \begin{example}
For a  $\T$-BP  with $m=r=2$  on $\mathbb{Z}^2$, one has that  $p_I^d(\T) = \frac{1}{8}$.
\end{example}
 
\subsection{Immunity for $\T$-BP on  social networks}

In order to understand the implications of $\T$-BP when modelling rumour spread in social networks, one needs to consider graphs $G$ which accurately represent society. For this, note that many   networks  have a power law degree distribution \cite{Onnela}, which means that the fraction of vertices in $G$ having degree $d$ is approximately $P(d)= d^{-\gamma}$ for some constant $\gamma \in \mathbb{R}$. For social networks particularly, $\gamma$ is often around between 2 and 3 (e.g. see \cite{Choromaski, Onnela}). However, although social networks are relatively random, graphs known as deterministic hierarchical networks have power law degree distributions while also having predetermined configurations, making them simpler to study. Thus, it becomes very interesting to investigate the $\T$-BP on a deterministic hierarchical network with $2\leq\gamma\leq3$.

\begin{figure}[h]
    \begin{center}
    \centering
    \includegraphics[width = 0.5\textwidth]{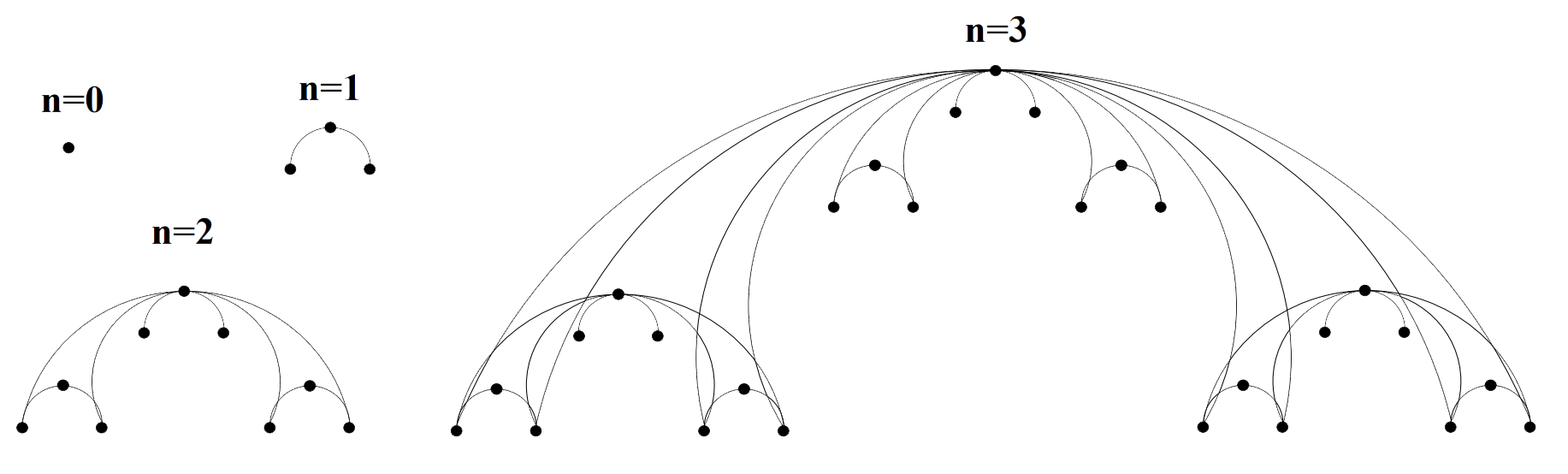}
    \caption{A  hierarchical network as constructed in  \cite{Hierarchical}.}
    \label{Hierarchy 2}
    \end{center}
\end{figure}

An example of a deterministic hierarchical network with $2\leq\gamma\leq3$   is  defined in \cite{Hierarchical}, and shown in Figure \ref{Hierarchy 2}. To construct it, start with one node, a root node. Add two more nodes and connect them to the root. At step $n$, add $3^{n-1}$ nodes each, identical to the figure in the previous iteration (step $n-1$) and connect the $2^n$ bottom nodes to the root.   The degree exponent for these graphs is $\gamma = 1 + \frac{\ln 3}{\ln 2}$.

At $n$ iterations there are $\left(\frac{2}{3}\right)3^{n-i}$ vertices with degree $2^{i+1}-2$. Hence,   substituting $k$ with $2^{i+1}-2$, one can see that there are $2(3^{n-\log_{2}(k+2)})$ vertices of degree $d$ at this step. Hence we have that 
at $n$ iterations, there are 
$2(3^{n-\log_{2}(d+2)})$
 vertices of degree $d$.
There are also $3^n$ vertices at $n$ iterations and thus the probability a vertex has degree $d$ at $n$ iterations is    $2(d+2)^{-\log_2 3}$. 
 
The maximum degree of any vertex on this graph is the degree of the root, which at step $n$ is given by $2^{n+1}-2$ \cite{Hierarchical}. Hence, this is the upper bound of the summation. To find $p_I$, or the expected fraction of initially immune vertices on this model with trust family $\T$, we need to find the probability of immunity for a vertex with $j$ neighbours and then multiply that by the probability that a vertex has $j$ neighbours. Finally, taking   this product across all possible $j$
one has that the expected fraction of immune vertices $p_I(G,\T)$ for this
hierarchical network is given by \begin{eqnarray}p_I(G,\T
)=\sum \limits_{d =
1}^{2^{n+1}-2} 2(d+2)^{-\log_2 3}p^d_I(\T).\end{eqnarray}.

      \section{$\T$-BP on random networks} \label{mas}

%
%
When studying a percolation model, one is particularly interested in the critical probability $p_c$ describing the initial probability of infection that would make at last half the graph infected by the end of the process. 
Since society can be modelled through random graphs, we shall dedicate the next section to the study of $\T$-BP on Erdos-Renyi graphs. In particular we study the following properties, for which the results are primarily analytical:
\begin{itemize}
    \item[(I)] The initial probability of infection $p$ and the critical probability of percolation $p_c$;
    \item[(II)] The fraction of the graph that is infected in $A_t$ for a  given time $t$, or in $A_\infty$.
    \end{itemize}
     Through these properties of the model  one can  understand how gossip spreads or how marketing models based on $\T$-BP behave.
    Note that the initial probability of infection $p$ determines the proportion of society that carries a gossip, or that have originally bought the objects for which the marketing campaign is being analysed.

 In what follows we shall consider the properties of  $\T$-BP on random networks
  in terms of the main variables of the $\T$-BP model: the time $t$; the number $m$ of labels a vertex may have,
 the number $r$ of different labels the set of infected neighbours must include in order for a vertex to be infected, and     the probability  $\rho$ of having an edge between two vertices.
Finally,  throughout this section, for any value requiring multiple trials, we run 30 trials and fix   $n:=|V(G)|=10000$ vertices.

\subsection{Variation of the model's density}

The percolation of the $\T$-BP model depends heavily on the probability  $\rho$ of having an edge between two vertices. Indeed, one can see that for larger values of $\rho$, very low initial probabilities of infection will lead to the whole network being percolated by the end of the process, this is, $A_\infty = G$. Moreover, there is a clear wall crossing phenomena for the fraction of the graph that is percolated by the end of the process: a slight shift in $p$ causes a large jump in the fraction percolated and also the probability of percolation. An example of this is given in Figure \ref{vary den}:

\begin{figure}[h]
     \centering
     \resizebox{0.45\textwidth}{!}{%
    \includegraphics{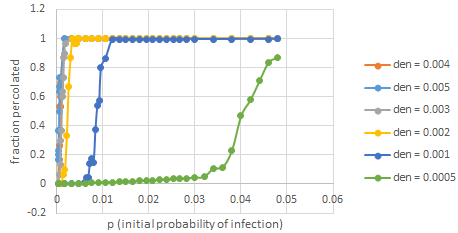}}
    \caption{Fraction of the graph percolated as the initial probability of infection $p$ varies, given various  values of the density  $\rho$, taking  the integers $n = 10000$, $m = 3$, $r = 2$.}
    \label{vary den}
\end{figure}

The above Figure \ref{vary den} shows the  relationship  between the fraction of a graph that would end up percolated and the initial probability of infection, $p$. We considered this for a $\T$-BP model with $m=3$ and $r=2$ on Erdos-Renyi graphs with various values for $\rho$. In particular, one can see that for densities $\rho\geq 0.001$, the model is very likely to completely percolate for any initial probability of infection $p> 0.01$.

\begin{example}Consider the setting of Example \ref{demo}, where a society of 10000 people who are Democrats, Republicans and Independents is modelled through  $\T$-BP   with $m=3,~r=2$ and $k_i\in \{0,1\}$. In this setting, from Figure \ref{vary den} one can see that if people are on average connected to 10 people or more (this is $\rho>0.001$), and more than 100 people initially believe certain gossip (this is $p>0.01$), then very quickly the whole society will carry that gossip. 
On the other hand, if the average person only exchanges gossip with 5 people or less (considering $\rho<0.0005$), even if 300 people initially carried the gossip, the whole society would likely not end up carrying the gossip.  In fact,  we would expect less than 2000 people carrying the gossip by the end  of the process (this is, the fraction of the graph percolated would be less than 0.2). 

   \end{example}

 Through our model, one can look at  individual curves in graphs as the ones in Figure \ref{vary den}, and determine the critical probability $p_c$ for some fixed value for $\rho$. 
\begin{figure}[h]
    \begin{center}
    \centering
        \resizebox{0.45\textwidth}{!}{
    \includegraphics{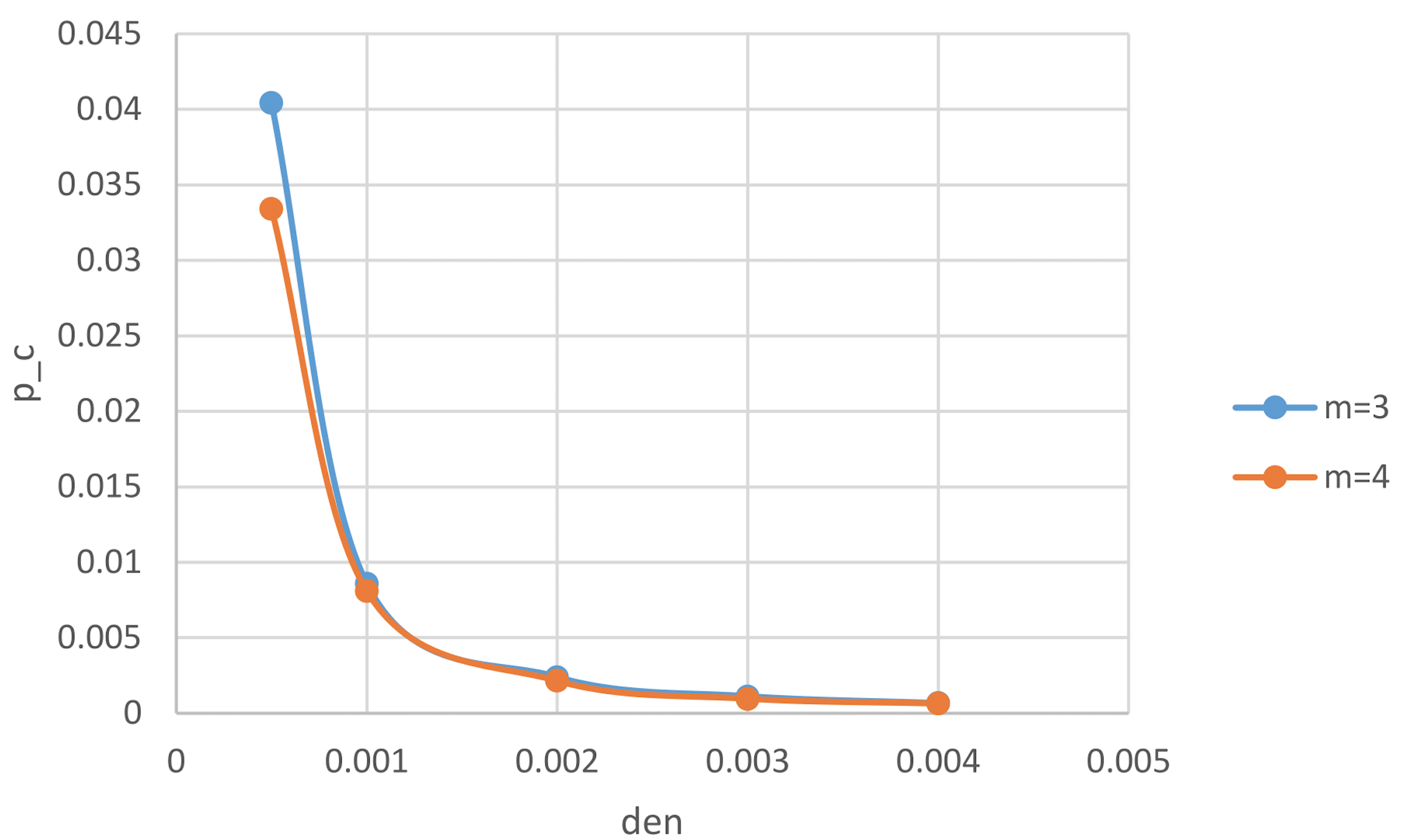}}
    \end{center}
    \vspace{-5mm}
    \caption{Variation of $p_c$ in terms of  the density $\rho$ for $m=3,4$, where $r=2$ and the number of vertices  is $n = 10000$.}
    \label{den vs p_c}
\end{figure}

\begin{figure}[h]
    \begin{center}
    \centering
        \resizebox{0.4\textwidth}{!}{
    \includegraphics{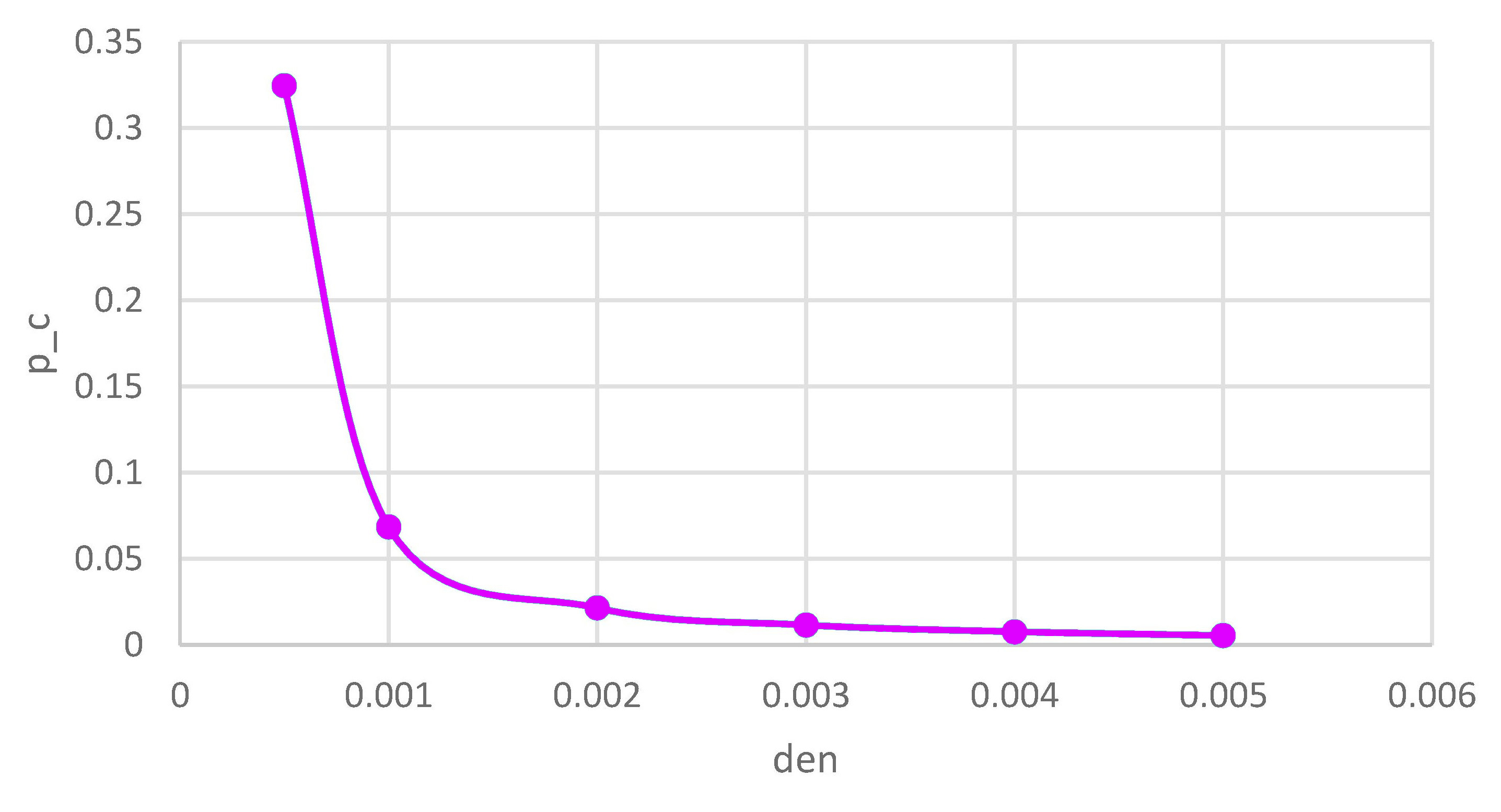}}
    \end{center}
    \vspace{-5mm}
    \caption{Variation of $p_c$ in terms of the density $\rho$ for $m=4$ and $r=3$, where the number of vertices  is $n = 10000$.}
    \label{den vs p_c_2}
\end{figure}

In  Figure \ref{den vs p_c} we compare the value of $p_c$ to the value of $\rho$ for a 2-neighbor $\T$-BP model with invariants $m=3,4$. As expected, one can see that the more labels one has (whilst requiring the same number of labels to be infected), the higher the critical probability is. In other words, the more different types of people a society has, the higher the initial probability of carrying a gossip needs to be in order for more than half the society to believe the gossip by the end of the process. 
Moreover, increasing the number of labels required makes the critical probability jump considerably. In particular, one should compare 
the blue curve in Figure \ref{den vs p_c} and the graph in Figure \ref{den vs p_c_2}.

\begin{example}\label{6}
In order to see the importance of the number of non-zero $k_i$ a model has, we shall study a variation of Example \ref{demo}. Consider a society of 10000 people who are either Democrats, Liberals, Independents or Politically Agnostic (hence, having $m=4$). Suppose that the average person exchanges gossips with 20 people (this is, such that  $\rho=0.002$). 

In this setting, if a person only passed a rumour if it had been heard by at least 2 different types of people, then at least 25 people would need to carry the gossip initially for it to spread to half the society or more (since $p_c\sim 0.0025$). In contrast, of one required a rumour to be heard from at least 3 different types of people before it could be spread, then the gossip would need to be initially believed by 250 people for it to spread to half the society or more -- this is,  an order of magnitude more people than if we required only two different types of carriers.  \end{example}

%
%

%


 \subsection{Variation of the model's type}\label{type}

We shall now consider the
 relationship  between the fraction of a graph that would end up percolated and the initial probability of infection $p$,   while varying the number  $r$ of required labels.  
As one would expect, a $\T$-BP model percolates much faster, and to a much larger proportion of the society, the larger the number $r$ of types required is, and this is illustrated in Figure \ref{p vs frac_perc as r varies} where we fixed $n, \rho$ and $m$, and vary the number of labels:

\begin{figure}[h]
    \centering
    \resizebox{0.5\textwidth}{!}{%
    \includegraphics{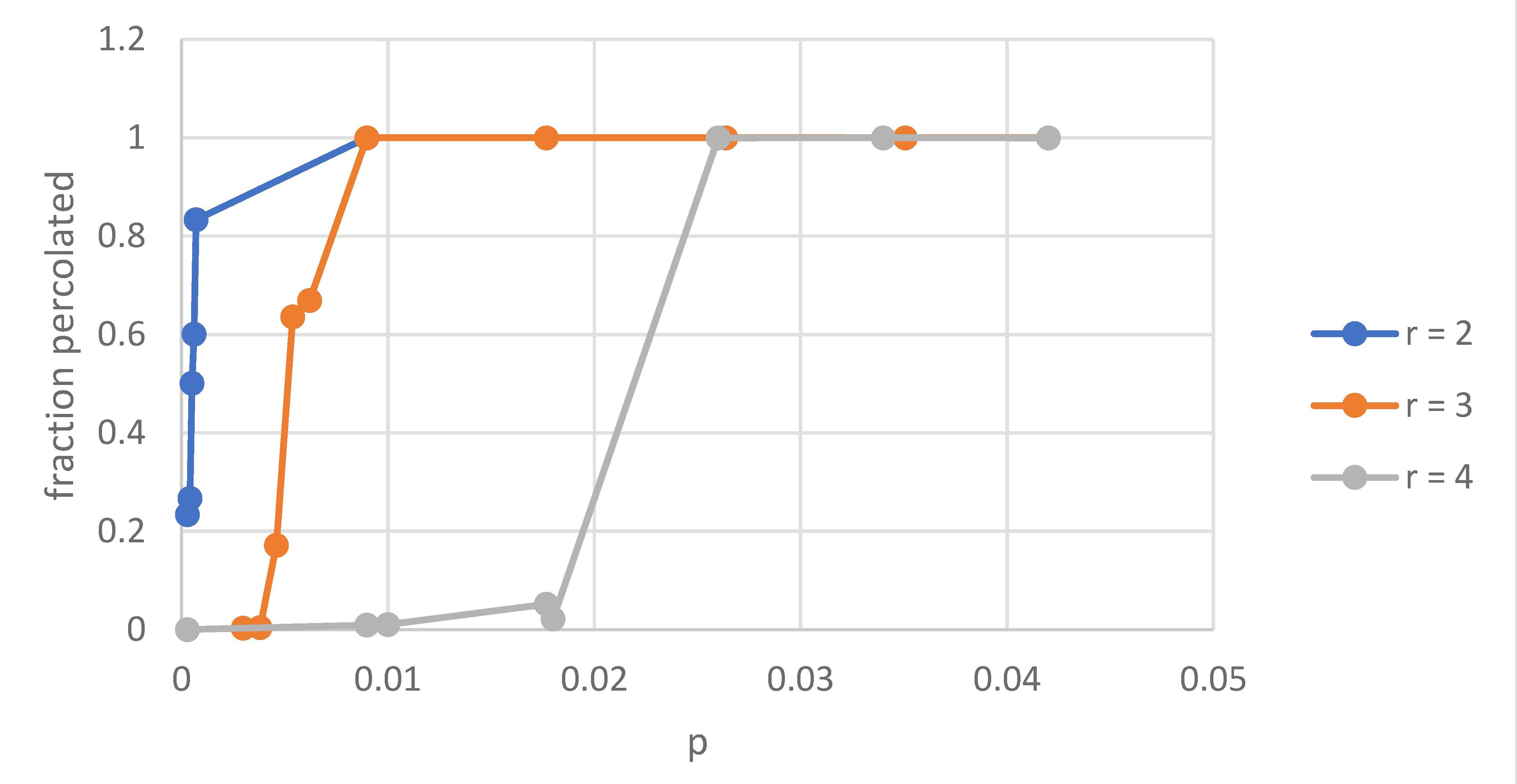}}\
    \caption{Fraction of the graph percolated in  terms of the variation of $p$ for varius $r$, where $n = 10000$, $m = 4$, and $\rho = 0.005$.}
    \label{p vs frac_perc as r varies}
\end{figure}

\begin{figure}[h]
    \begin{center}
     \centering
        \resizebox{0.5\textwidth}{!}{
    \includegraphics{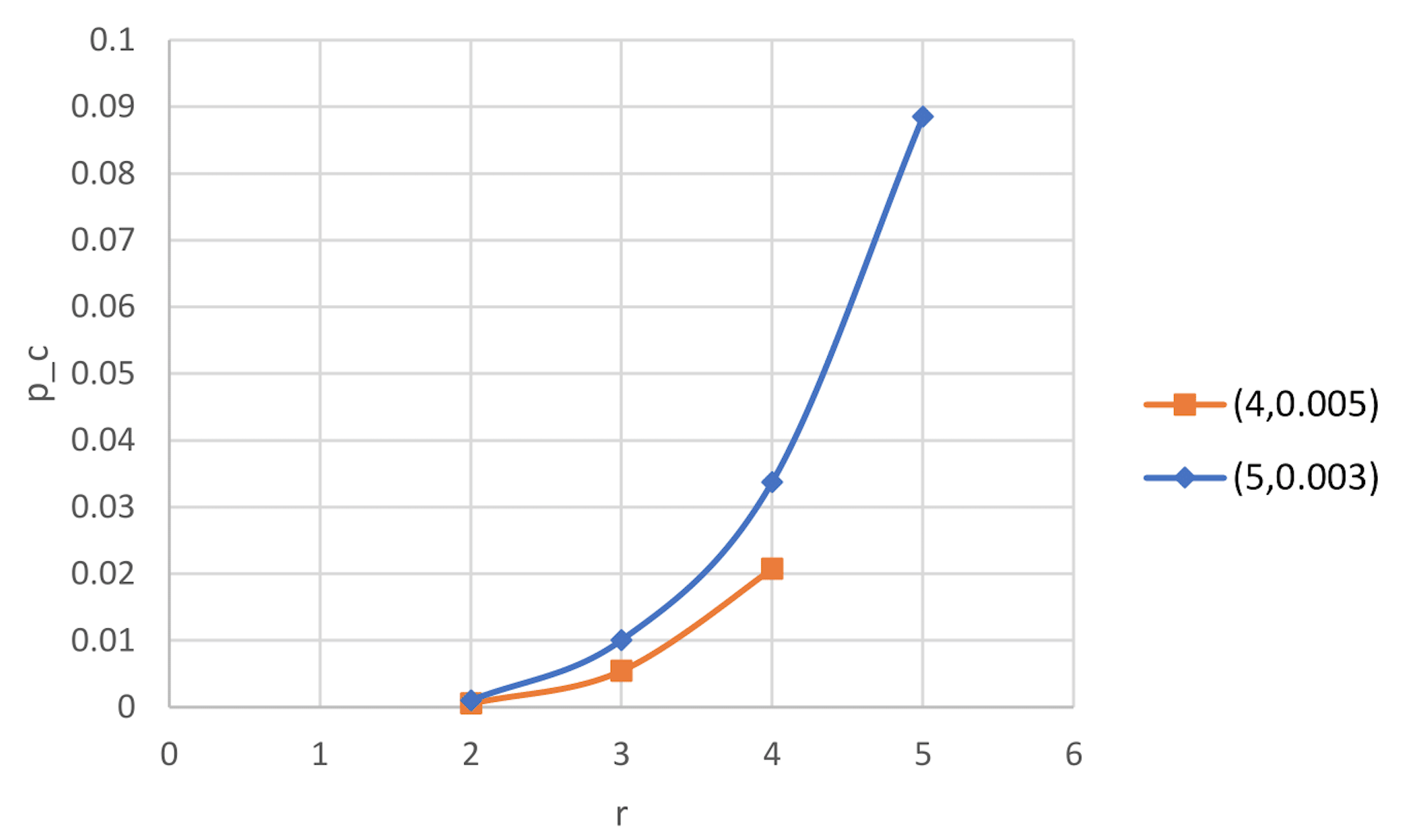}}
    \end{center}
     \caption{Variation of  $p_c$ with respect to $r$ for $(m,\rho)=(4,0.005)$ and $(m,\rho)=(5,0.003)$,  for $|V(G)|=10000$.}
    \label{r vs p_c}
 \end{figure}

\begin{example}
Similar to Example \ref{6}, consider  a society of 10000 people who are either Democrats, Liberals, Independents or Politically Agnostic (hence, having $m=4$), but suppose now that   the average person exchanges gossips with 50 people (this is, such that  $\rho=0.005$). Moreover, suppose that initially only 200 people believe certain gossip ($p=0.02$). By requiring someone to believe the gossip only if it is heard by every type of people,  the gossip would not spread over more than 25$\%$ of society. If instead one only required any smaller number of different types of people, the gossip would always spread over the whole society by the end of the process.

\end{example}
In Figure \ref{r vs p_c}   we plot $r$ against $p_c$, for various pairs of parameters $(m,\rho)$. As in the previous case, $p_c(r)$ appears to fit a power law curve again, albeit more weakly in some cases.

\pagebreak
 Numerically, one can see that for a fixed $m$ the probability of percolation $p_c(r,\rho)$ seems to obey a power law in  both variables $\rho$ and $r$, suggesting that 
$$p_c(r,\rho) = cf(m){r}^{e_1}{\rho}^{-{e_2}},$$ where $c, e_1, e_2$ are all positive constants. Moreover, varying $m$ also affects $p_c$, which is why the $f(m)$ component is present.

We shall finally consider  the average growth curve for fixed values of total types $m$, number of types required by the model $r$ and density $\rho$. In terms of the initial probability of infection, we can see the growth in Figure \ref{avg growth} where we considered the same density as in Figure \ref{p vs frac_perc as r varies} and Figure \ref{r vs p_c} to allow for comparison:

 \begin{figure}[h]
    \begin{center}
    \centering
        \resizebox{0.45\textwidth}{!}{%
    \includegraphics{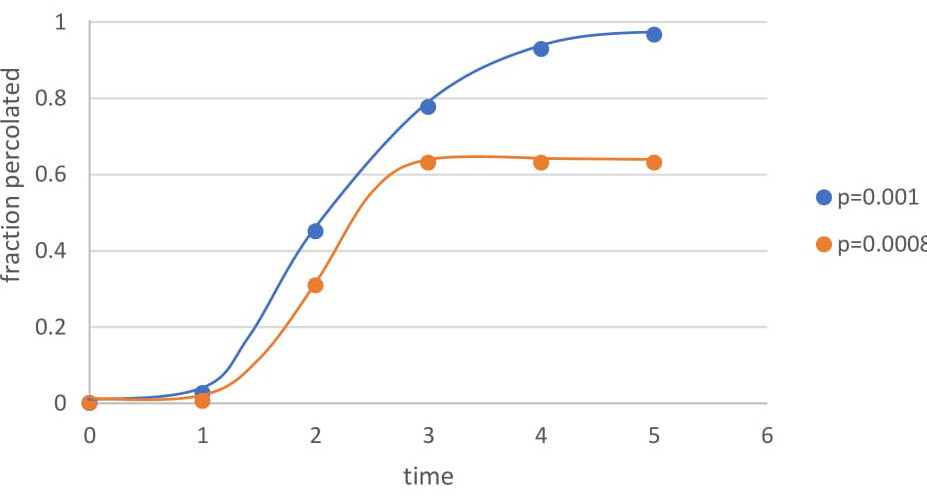}}
    \caption{Average growth curve (comparing the $t$ to $\text{frac}_\text{perc}$) for graphs with $m= 3$, $r = 2$, $n = 10000$, and  $\rho = 0.005$.}
    \label{avg growth}
    \end{center}
\end{figure}

\begin{example}
One can see the relevance of Figure \ref{avg growth} through the setting of Example \ref{demo}. Indeed, consider a society of Democrats, Republicans, and Independents where one believes gossip only if at least two people from different parties tells it. Then, if originally only 8 people believe a gossip, by the end of the $\T$-BP process more than 30$\%$ of the society will not believe the gossip. On the other hand, if two more people believe the gossip originally (hence, having $p=0.001$), by the end of the percolation process all members of the society will believe it. 
\end{example}

\section{Conclusion}\label{last}
Bootstrap percolation has been used for years to model various percolation processes, with applications spanning from epidemiology to rumour spreading.
The $\T$-BP model has been developed to add an extra layer to the current forms of bootstrap percolation, as the vertices of the graphs on which the infection spreads are now labeled. One should note that whilst many models have been developed to understand rumour spread (e.g. see \cite{calio,Rum1,Rum4,Rum2,lind,Rum3,Mas}), these models are very different to $\T$-BP, and non of them consider multi-type percolations. 

The  \textit{\textbf{$\T$-Bootstrap percolation}} ($\T$-BP model) was originally created to represent the spread of information based on the basic human instinct to trust information more if it comes from a variety of sources.
This multi-type Bootstrap percolation model  can be used to deduce interesting properties of social networks and their behaviour within different contexts (e.g., rumour spread,  marketing, infection spread, etc.). Whilst it is interesting to analyse the model on arbitrary random networks, it is of particular interest to consider random networks with plausible sizes and densities that accurately reproduce society (e.g. see \cite{Hierarchical}, \cite{Otte}, \cite{Opsahl}, \cite{fox}, \cite{Chakrabarti}).

In this paper we considered the hierarchical networks introduced in  \cite{Hierarchical} to described how a rumour or disease spread through a $\T$-BP model within the network of density between 2 and 3. In particular, we can see the following remarkable behaviours appear: 

\begin{itemize}
\item {\bf Delay in the spread}.
Infections spread much faster through the classical $r$-neighbour bootstrap percolation than via the $\T$-BP (e.g.~see Figure \ref{TMBP vs classical});
  
\item { \bf Containment of  the infection}. An infection  spreads across a greater percentage of the population  in the classical $r$-neighbour bootstrap percolation than via the $\T$-BP (e.g.~see Figure \ref{TMBP vs classical});

\item  { \bf Trust vectors}. By requiring higher values of $r$, the percolation of the model is delayed (see for example Figure \ref{p vs frac_perc as r varies}).  
\end{itemize}

 In particular, we can see how effective the $\T$-BP is in hindering the spread of fake news, and how by requiring higher levels of trust (higher values of $r$), gossip would spread more slowly and to a lesser amount of people. 
 
%

Interestingly within the $\T$-BP models, some vertices may be immune, and in this paper we studied the probability of a vertex being immune. The immunity of a vertex may be determined before the initial infection of the graph even occurs, as it is based upon a vertex's degree and the update rule for the graph. The probability $p_I$ of immunity was presented here for  the  $\T$-BP  as $r$-bootstrap percolation, the simplest form of $\T$-BP with only one trust vector, and $\T$-BP on the deterministic hierarchical graphs. Moreover, we were able to estimate  the expected number of vertices which will be immune on any particular graph with these models. We also offer a rather loose lower bound for $p_c$, the critical probability of infection, based on $p_I$. Finally, we concluded our investigation by looking to random graphs, as these better represent the irregularities of society, and deriving analytical results on these graphs by running the $\T$-BP model on them computationally. 

We expect the study of  $\T$-BP models to be of particular interest from many different perspectives, and we shall conclude this paper mentioning a few of these lines of research which we plan to investigate in the future:

\begin{itemize} 
\item {\bf Deterministic hierarchical graphs}. These graphs seem promising for applications in sociology and marketing, as society has the same power law degree distribution as these graphs do, while their deterministic nature makes them simpler to study (as opposed to Erdos-Renyi graphs, which add another element of randomness to the investigation, thus making it more complex). Progressing further, we may wish to take our study to random graphs with this power law degree distribution, and eventually make the model stochastic by introducing probability of the infection spreading from one vertex to another.

\item {\bf Echo chambers}. Social media networks and search engines keep track of news a user and his/her friends respond positively to, and then use this information to suggest future articles and advertisements. However, when it comes to news content and discussion of the news, this means one will increasingly only see material that is in line with one's stated interests. This worsens issues of polarisation and group-think. To combat this, we could apply the $\T$-BP to the algorithms which suggest the news users are shown -- this would prevent the existence of echo chambers by forcing people to see information which might cater to multiple sides of the political spectrum.  

\item {\bf Genetic diseases}. One could consider  a graph $G$ with vertices $1$ and $2$, where each label represents a possible sex. Then, in order for a vertex to be infected, it must be infected by two vertices above it of different sex. This represents how dominant X-linked genetic diseases are only passed on if both the male and female parent have the disease, and would be modelled by a $\T$-BP model with $r=2$. For recessive X-linked diseases, the model would need to have 3 labels -- an infected female, a female carrier, and an infected male, and would need to be stochastic, as the probability of infection from a female carrier and infected male would only be $\frac{1}{2}$.

\end{itemize}

In order to study the propagation of an infection or a rumour in a random society, one would like to consider graphs with realistic densities. For this, recall that Dunbar found in  \cite{dunbar} that  the expected number of acquaintances any individual may have is 150. Hence, by consider $\rho=0.015$ with our current model, one can obtain representations of society through Erdos-Renyi graphs. 

In the above setting, one has  that a potential application for the $\T$-BP is in news-suggesting algorithms on social media:   using the $\T$-BP as opposed to current algorithms could hinder the spread of fake news. Indeed, one can see the difference of the two models  by comparing the growth of infection on the $\T$-BP as opposed to on classical $r$-neighbour bootstrap percolation, for current news-suggesting algorithms work very similarly to the latter model. As an example, we compare the average growth on 3-neighbour 5-state $\T$-BP and on 3-neighbour classical bootstrap percolation in Figure \ref{TMBP vs classical}:

 \begin{figure}[h]
    \centering
        \resizebox{0.5\textwidth}{!}{%
    \includegraphics{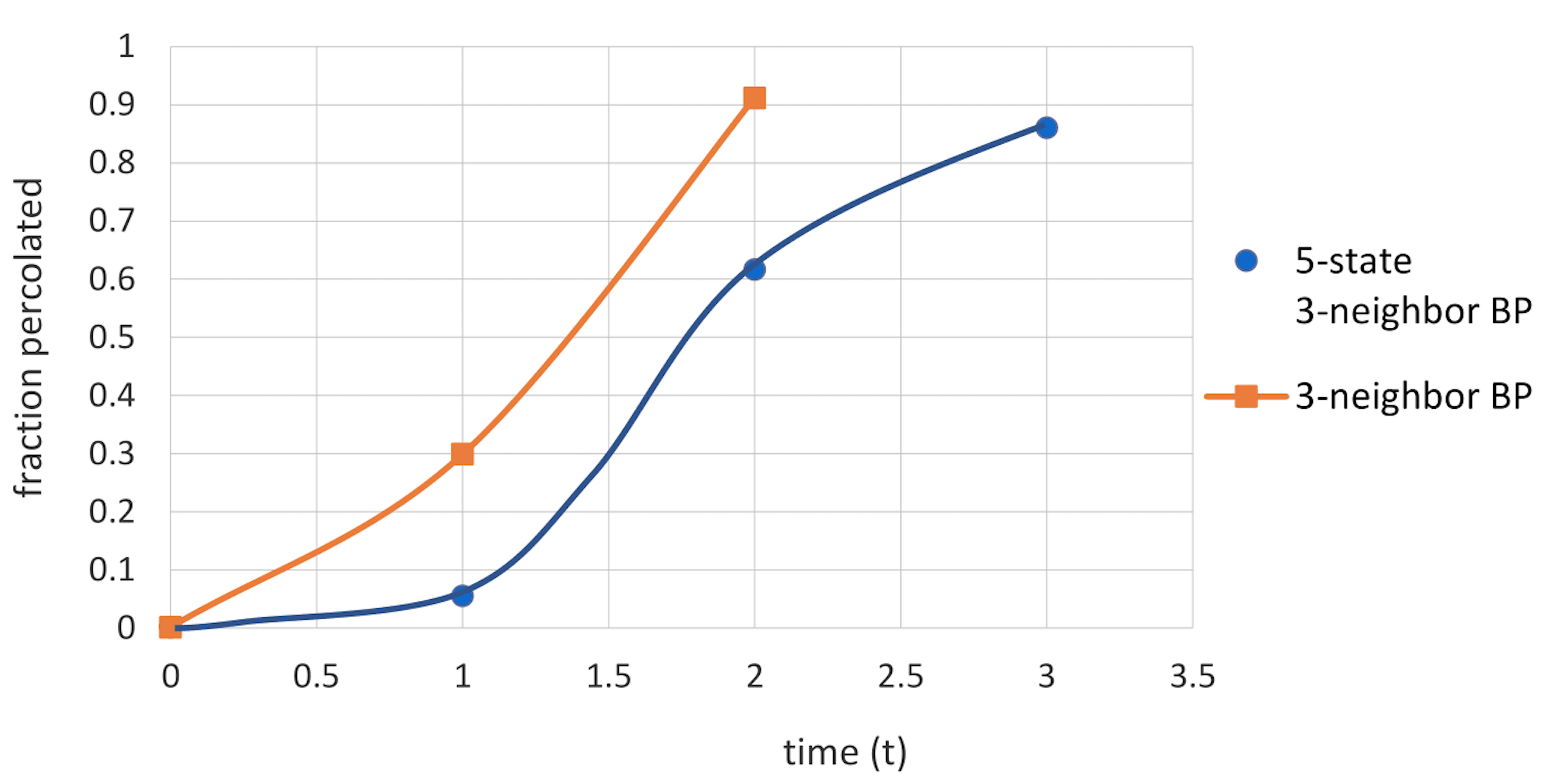}}
    \caption{Average growth curve   for graphs with $\rho = 0.015$, $n = 10000$, $p = 0.0015$, and $(m,r) = (5,3)$ (for the $\T$-BP) and then $r=3$ (for the classical model).}
    \label{TMBP vs classical}
\end{figure}
The process for the $\T$-BP lasts an entire time step longer, and the growth is more gradual and a lower fraction of the society is infected by the end. In other words, the $\T$-BP makes biased news spread more slowly and to a lesser degree, indicating potential applications in replacing current news-suggesting algorithms on these social media platforms. Additionally, note the usage of $\rho=0.015$ in these trials, as they are being conducted specifically to determine the potential of the $\T$-BP on real social networks.

We shall conclude by mentioning that  having knowledge of the value of $p_c$ for any given social network is important for numerous reasons. For instance, it is incredibly useful for marketers and journalists to be aware of, since the number  $p_c$ gives the minimum number of people they must directly target with news of some product (through advertisements, free samples, etc.) or perhaps some scandal so that the probability of percolation, which is conceptually analogous to the probability of this information going viral, is $\frac{1}{2}$. Thus, developing a strong general formula for $p_c$ would be very useful.

 \bigskip

 \noindent {\bf Acknowledgments:} The authors are thankful to PRIMES-MIT for their support, as well as to David Conlon and Dhruv Mubayi for useful conversations, and to James Unwin for useful comments on a draft of the paper. The work of L.P. Schaposnik is partially supported by grants NSF DMS 1509693 and NSF CAREER Award DMS 1749013, and she is thankful to the Simons Center for Geometry and Physics for the hospitality during part of the preparation of the manuscript.

\pagebreak
 
 \begin{appendix}
 \section{The probability of immunity I}\label{AA}
 \noindent{\bf Proposition}.  {\it Consider $\T$-BP on a graph $G$, and let each vector component $k_j\in \{0,1\}$. Then, the probability   that a vertex $v$ with $|N(v)|=d$ has $D_v\leq r-1$   is \begin{eqnarray}p^d_I(m,r):=P(D_v\leq r-1)=  \frac{ \sum\limits_{j = 1}^{r-1} \left[{m\choose j}(j!)\stirling{d}{j}\right]
}{m^d},\nonumber
\label{D}\end{eqnarray}
where $\stirling{d}{j}$ is the Stirling number of the second kind. }
\smallskip
\begin{proof}
We shall prove the above statement through the principle of inclusion and exclusion. In order to do this, we shall first calculate the probabilities 
$P(D_v = i)  ~ {~\rm for~}~ i=1, \ldots, r,$
and use this to calculate $p^d_I(m,r)$.

Given a fixed integer $n$ such that $1\leq n\leq r-1$,  in order to understand  $P(D_v = n)$  note that there are ${m \choose n}$ ways to choose the $n$ acceptable labels that the neighbours may have, or in other words, ways to choose a set $N\subset \{1,\ldots, m\}$ with $|N|=n$.
Once these labels are chosen, the number of onto functions 
 such that $f(N(v))=N$ is given by
\begin{eqnarray}
\sum\limits_{i=0}^{n} (-1)^{n-i}{k \choose i}(n-i)^d.
\end{eqnarray}
To see this, begin by noting that each of the $d$ adjacent vertices in $N(v)$ has $n$ possible values for its label, so there are $n^d$   functions from $N(v)$ to $N$. However, not all functions will be different. To visualise this, consider   a Venn diagram where each   circle $A_s$ of the Venn diagram corresponds to a label $s$, with $1 \leq s \leq m$, and is defined as $A_s:= \{f:N(v)\rightarrow N~:~s\not\in N\}.$
For instance, every object in $A_{1}$ will be a possible configuration of neighbours of $v$ such that none of them are of label 1. 
 
All functions which are not surjective to $N$ must appear in some set $A_i$ for $i\in N$. Therefore, the number of onto functions $f:N(V)\rightarrow N$ is given by \begin{eqnarray}n^d - |\bigcup\limits_{i \in N} A_{i}|.\label{all}\end{eqnarray}
In order to calculate \eqref{all}, recall that by the Principle of Inclusion and Exclusion,
\begin{eqnarray}|\bigcup\limits_{i=1}^{\alpha}A_{i}| = \sum\limits_{i=1}^{\alpha}|A_i| - \sum\limits_{1 \leq i < j \leq \alpha}|A_i \cap A_j| +\\
 \sum\limits_{1 \leq i < j < k \leq \alpha} |A_i \cap A_j \cap A_k| + ... + (-1)^{\alpha-1}|\bigcap\limits_{i=1}^{\alpha}A_i|.\end{eqnarray}
Thus, to find $|\bigcup\limits_{i \in N} A_{i}|$ one needs to consider $\sum |A_{x_1} \cap ... \cap A_{x_j}|$ for all $j$ such that $1 \leq j \leq n$, and $x_i\in N$, which is given by
\begin{eqnarray} \sum |A_{x_1} \cap ... \cap A_{x_j}| ={n \choose j}(n-j)^d,
\end{eqnarray}
since there are ${k \choose j}$ ways to choose the $j$ labels out of $N$ that the neighbours cannot occupy, and so there are $n-j$ options for every neighbour's label. Then, one has that  
\begin{eqnarray}
|\bigcup\limits_{i=1}^{n}A_{i}| = \sum\limits_{i=1}^{n} (-1)^{n-i+1}{n \choose i}(n-i)^d
\end{eqnarray}
and thus the number in \eqref{all} of surjective functions \linebreak $f: N(v) \rightarrow N$ is
 \begin{eqnarray}
\sum\limits_{i=0}^{n} (-1)^{n-i+1}{n \choose i}(n-i)^d.
\label{funct}
\end{eqnarray}
From the above, one can calculate the number of different label assignments that the vertices $N(v)$ can have,  where $D_v = j$: there are $m \choose j$ possibilities for   $j$ labels, which  multiplied by \eqref{funct} leads to \begin{eqnarray}
 {m \choose j}\sum\limits_{i=0}^{j} (-1)^{j-i+1}{j \choose i}(j-i)^d.
 \end{eqnarray}
Equivalently, we can write this as ${m\choose j}(j!)\stirling{d}{j}$ 
where $\stirling{d}{j}$ is the Stirling number of the second kind. Hence, the number of different label assignments to the graph $G$ for which the diversity $D_v\leq r-1$ is   \begin{eqnarray}
 \sum\limits_{j = 1}^{r-1} \left[{m\choose j}(j!)\stirling{d}{j}\right].
 \end{eqnarray}
Recalling that there are $m^d$ possible functions $f: N(v)\rightarrow \{1,\ldots,m\}$,   the probability of a vertex having diversity $D_v\leq r-1$ is given by 
 \begin{eqnarray}
 P(D_v\leq r-1)= \frac{ \sum\limits_{j = 1}^{r-1} \left[{m\choose j}(j!)\stirling{d}{j}\right]
}{m^d},
\end{eqnarray}
which concludes the proof.
\end{proof}

\section{The probability of immunity II}\label{BB}

\noindent{\bf Proposition}. {\it In a $\T$-BP  with $k = (k_1, ..., k_m)$, the probability of immunity $p^d_I(k)$ for a vertex $v$ with $|N(v)|=d$ is 
\begin{tiny}
\begin{eqnarray}
    p^d_I&(&K) =
        \notag 1- \sum \limits_{x_m = k_m}^{d-(\sum_{l=1}^{m-1}k_l)}{d\choose x_m}\left(\frac{1}{m}\right)^{x_m} \\
        \notag   &\bigg[&\sum \limits_{x_{m-1} = k_{m-1}}^{d-x_m-(\sum_{l=1}^{m-1}k_l)}{d-x_m\choose x_{m-1}}\left(\frac{1}{m}\right)^{x_{m-2}} \bigg[ \cdots \\ 
        \notag  &\bigg[&\sum \limits_{x_2 = k_2}^{d-(\sum_{l=3}^{m} x_l)-k_1}{d-(\sum_{l=3}^{m} x_l)\choose x_2}\left(\frac{1}{m}\right)^{x_2}\left(\frac{1}{m}\right)^{d-(\sum_{l=2}^{m} x_l)} \bigg] \bigg] ... \bigg].
\label{second}
\end{eqnarray}
\end{tiny}}
\smallskip
\begin{proof}
Let  $f(a,b,c) $ be the function defined as
\begin{tiny}\begin{eqnarray}
    f&(&a,b,c) :=
        \notag \sum \limits_{x_i = b_i}^{a-(\sum_{l=1}^{i-1}b_l)}{a\choose x_i}\left(\frac{1}{c}\right)^{x_i} \\
        \notag ~&\bigg[&\sum \limits_{x_{i-1} = b_{i-1}}^{a-x_i-(\sum_{l=1}^{i-1}k_l)}{a-x_i\choose x_{i-1}}\left(\frac{1}{c}\right)^{x_{i-1}} \cdots \\
        \notag  &\bigg[& \sum \limits_{x_2 = b_2}^{a-(\sum_{l=3}^{i} x_l)-b_1}{a-(x_i+...+x_3)\choose x_2}\left(\frac{1}{c}\right)^{x_2}\left(\frac{1}{c}\right)^{a-(\sum_{l=2}^{i} x_l)} \bigg] ... \bigg]  
\end{eqnarray} \end{tiny}where $
    b = \{b_1, b_2, \cdots, b_i\}
$ and $a,c \in \mathbb{N}$.
Then, one can see that $p_I = 1 - f(d,k,m)$ for $k = \{k_1, k_2, \cdots, k_m\}$.
Indeed, this can be proven with an inductive argument on the number of available labels $m$. 
 For $m = 2$, the vector $k = \{k_1,k_2\}$ satisfies $k_1 + k_2 \leq d$. We must find the probability of assigning $d$ vertices to one of 2 labels such that there are at least $k_1$ vertices of label 1 and $k_2$ of label 2--this holds when a vertex is not immune, so we must then subtract this from 1. This is equivalent to saying there may be $x_1$  vertices of label 1 such that $k_1 \leq x_1 \leq d-k_2$, and all other vertices of label 2. Note that the probability that there are $x_1$ vertices of label 1 is:
\begin{eqnarray}
    {d \choose x_1}\left(\frac{1}{2}\right)^{x_1}\left(\frac{1}{2}\right)^{d-x_1}.
\end{eqnarray}
Summing over all possible values for $x_1$ varying from $k_1$ to $d-k_2$, the overall $p_d$ for  $m=2$ and $k = \{k_1,k_2\}$ is found to be:
\begin{eqnarray}
    \sum \limits_{x_1 = k_1}^{d-k_2} {d \choose x_1}\left(\frac{1}{2}\right)^{x_1}\left(\frac{1}{2}\right)^{d-x_1}.
\end{eqnarray}
Note that this equals $f(d,\{k_1,k_2\},2)$. Then, the probability of immunity would be
\begin{eqnarray}
    1 - \sum \limits_{x_1 = k_1}^{d-k_2} {d \choose x_1}\left(\frac{1}{2}\right)^{x_1}\left(\frac{1}{2}\right)^{d-x_1}.
\end{eqnarray}
Now, move on to $m=3$ with $k = \{k_1,k_2,k_3\}$. We must find the probability of assigning $d$ vertices to one of 3 labels such that there are at least $k_1$ vertices of label 1, $k_2$ of label 2, and $k_3$ vertices of label 3. We can approach this with casework.

First, note that there may be $x_1$ vertices of label 1 such that $k_1 \leq x_1 \leq d-(k_2+k_3)$. Then, take cases based on the value of $x_1$. Note that given $x_1$, there are $d-x_1$ remaining vertices to consider, with 2 possible labels to assign them to. Now this question is almost the same as the one with $m=2$, with the only difference being that the probability a vertex occupies one of these two labels (2 or 3) is not $\frac{1}{2}$, but $\frac{1}{3}$. So given $x_1$, the probability that the number of vertices of label 2 is greater than $k_2$ and the number of vertices of label 3 is greater than $k_3$ is just $f(d-x_1,\{k_2,k_3\},3)$. This means that given $x_1$, the probability that the vertex is initially not immune (do not forget that we are using complementary counting) is:
\begin{eqnarray}
    {d \choose x_1}\left(\frac{1}{3}\right)^{x_1}f(d-x_1,\{k_2,k_3\},3).
\end{eqnarray}
Now, as $x_1$ can go from $k_1$ to $d-(k_2+k_3)$, $p_I$ for $m=3$ is:
\begin{eqnarray}
    1 - \sum \limits_{x_1 = k_1}^{d-(k_2+k_3)} {d \choose x_1}\left(\frac{1}{3}\right)^{x_1}f(d-x_1,\{k_2,k_3\},3)
\end{eqnarray}.
Note that this equals $f(d,\{k_1,k_2,k_3\},3)$. 
Therefore, 
\begin{eqnarray}\nonumber
    p^d_I&(&\{k_1,k_2,k_3\}) =
       1 - f(d,\{k_1,k_2,k_3\},3)\\&=&
            1 - \sum \limits_{x_1 = k_1}^{d-(k_2+k_3)} {d \choose x_1}\left(\frac{1}{3}\right)^{x_1}f(d-x_1,\{k_2,k_3\},3),\nonumber
\end{eqnarray}
which provides the intuition for the inductive step: we must prove that $$p^d_I(\{k_1,...,k_m\}) = 1 - f(d,\{k_1,...,k_m\},m),$$
assuming that there exists an $i$ such that 
$p^d_I(\{k_1,...,k_i\}) = 1 - f(d,\{k_1,...,k_i\},i).$

In order to prove the statement by induction, consider $m=i+1$, and let $k=\{k_1,k_2, \cdots, k_{i+1}\}$. Using casework as we did for the example where $m=3$, let   the number of vertices of label $1$ be $x_{1}\in \{k_1,\ldots, d - \sum \limits_{j=2}^{i+1} k_j\}.$ Then, there are $d-x_1$ vertices  which must have labels in $[i+1]/\{1\}$. By the inductive assumption, the probability that they have at least $k_j$ vertices of label $j$ is $f(d,\{k_2,...,k_{i+1}\},i)$. However, note that we must replace the $\frac{1}{i}$ terms in this formula with $\frac{1}{i+1}$ because the probability any individual vertex has a specific label is now $\frac{1}{i+1}$, meaning that it is actually $f(d,\{k_2,...,k_{i+1}\},i+1)$. Also, the probability that there are $x_1$ vertices of label 1 is ${d \choose x_1}(\frac{1}{i+1})^{x_1}$, so multiplying these two terms one finds that the probability that a vertex is initially \textit{not} immune given that it has $x_1$ neighbours of label 1 is :
\begin{eqnarray}
     {d \choose x_1}\left(\frac{1}{i+1}\right)^{x_1}f(d,\{k_2,...,k_{i+1}\},i+1).
\end{eqnarray}

Since $x_1$ can be anything from $k_1$ to $d - \sum \limits_{j=2}^{i+1} k_j$, summing over the possible values of $x_1$ and subtracting this summation from 1 leads to the probability 
\begin{small}
\begin{eqnarray}
     &p_I^d&(\{k_1, k_2, \cdots, k_{i+1}\}) =\\
      &1 & - \sum \limits_{x_1 = k_1}^{d - (\sum_{l=2}^{i+1}k_l
      )} {d \choose x_1}{\left(\frac{1}{i+1}\right)^{x_1}f(d,\{k_2,...,k_{i+1}\},i+1)}.\nonumber 
\end{eqnarray}
\end{small}
Finally, note that by definition  this is simply $1 - f(d,\{k_1,...,k_{i+1}\},i+1)$. This concludes the inductive step, and so we have proven that $$p_I^d( \{k_1, k_2, \cdots, k_{i+1}) = 1 - f(d,\{k_1,...,k_{i+1}\},i+1),$$ finalising  the proof.
\end{proof}

\end{appendix}
  \begin{small}
  
 \end{small}

\end{document}